\documentclass[useAMS,usenatbib]{newmn2e}
\usepackage{times}

\usepackage[dvips]{graphicx}
\newenvironment{MyCaption}{\footnotesize}

\newcommand{\MySub}[1]{$_{\ensuremath{\mbox{\footnotesize{#1}}}}$}

\title[Figure
  Rotation of Dark Matter Halos]{Figure Rotation of Dark Halos in Cold Dark Matter Simulations}

\author[S.E.Bryan and C.M.Cress]{ S. E. Bryan$^{1}$ and C. M. Cress$^{1,2}$\thanks{Email: cressc@ukzn.ac.za} \\ %\footnotemark[1]\\
$^{1}$Astrophysics and Cosmology Research Unit, University of KwaZulu-Natal, Westville, South Africa \\
$^{2}$Physics Dept., University of the Western Cape, Private Bag X17,
7535 Bellville, South Africa}

\begin{document}

%\pagerange{\pageref{firstpage}--\pageref{lastpage}} \pubyear{2002}

\maketitle

\label{firstpage}

\begin{abstract}
We investigate the figure rotation of dark matter halos identified in
$\Lambda$CDM
simulations. We find that when strict criteria are used to select suitable
halos for study, 5 of the 222 halos identified in our z=0 simulation output
undergo coherent figure rotation over a 5 $h^{-1}$\,Gyr period. We discuss the
effects of varying the selection criteria and find that pattern speeds for a
much larger fraction of the halos can be measured when the criteria
are relaxed. Pattern speeds measured over a 1 $h^{-1}$\,Gyr period follow a log-normal
distribution, centred at $\Omega_p$ = 0.25 $h$ rad Gyr$^{-1}$ with a maximum value of
0.94 $h$ rad Gyr$^{-1}$. Over a 5 $h^{-1}$\,Gyr period, the average pattern speed
of a halo is about 0.1 $h$ rad Gyr$^{-1}$ and the largest pattern speed found is
0.24 $h$ rad Gyr$^{-1}$. Less than half of the selected halos showed alignment between their figure rotation axis and minor axis, the exact fraction being somewhat dependent on how one defines a halo. While the pattern speeds observed are lower than those
generally thought capable of causing spiral structure, we note that
coherent figure rotation is found over very long periods and argue that further
simulations would be required before strong conclusions about spiral
structure in all galaxies could be drawn. We find no correlation between
halo properties such as total mass and the pattern speed. 

\end{abstract}


\section{Introduction}

Initially, calculations of stellar orbits in galaxies ignored the dark matter
halo completely.  Then, as observational evidence for dark matter halos
increased \citep{bib:dmhalos}, orbit calculations involved modelling the halos
as stationary spherical objects.  Further investigation into dark matter halos
indicated that many halos were in fact triaxial
\citep{bib:dubinskicarlberg,bib:warren92} and not spherical.  This was taken
into consideration in, for example, \cite{bib:Schwarzschild93} but orbits were
still modelled within stationary dark matter halos and these calculations fail to take
into account the effect of figure rotation. \cite{bib:bekki02} suggest that
figure rotation of the dark halo would play an important role in the formation
and evolution of the embedded galaxies.  In particular, they suggest that
figure rotation may influence the formation of stellar bars as well as spiral
arms and warps.  They also suggest that figure rotation could trigger
star-bursts in galaxies, even at high redshifts.  Since orbits passing near
the centre of galaxies are affected, figure rotation could also have
consequences for black hole growth in galaxies \citep{bib:blackholes} and the
so-called `cusp/core problem' \citep{bib:cuspycore}.  Observations of such
effects could be used to constrain current galaxy formation theories.   
\ \\

Initial work on the subject of figure rotation was done by Dubinski in 1992.
He found that halos in simulations which included tidal effects experienced a
noticeable rotation.  The halos were found to have pattern speeds ranging from 0.1 to 1.6 radians per Gyr.  (These units are
essentially equivalent to km s$^{-1}$ kpc$^{-1}$).  The small sample
size and somewhat artificial implementation of the tidal field, meant further
studies in this area were necessary. \citet{bib:pfitzner99}, noticed that
a significant number of triaxial halos produced in cold dark matter
simulations tended to rotate steadily around their minor axes.  One such halo
was extracted from the simulation at late times and was allowed to evolve in
isolation over a 5 Gyr period.  During this time the halo was found to
rotate as a solid body with constant speed of about
1.1 radians per Gyr.  In 1999, \citeauthor{bib:bureau99} proposed that slow figure rotation of a surrounding
triaxial dark halo may be responsible for creating the spiral structure seen
in NGC 2915 and encouraged further studies of gas disks contained within the
potentials of rotating triaxial dark halos.   \cite{bib:bekki02} continued
this work, using numerical simulations.  They showed that a `slowly' rotating
triaxial dark halo could cause large spiral arms in an extended gas disk, and
that the resulting structure is strongly dependent on the pattern speed of the
halo.  \cite{bib:masset03} further explored the effects of a rotating triaxial
dark halo proposed by \cite{bib:bureau99}, using hydrodynamical simulations
and concluded that, while a rotating triaxial halo could be responsible for
causing spiral structure, the required pattern speed was prohibitively large -
between 5.5 and 6.5 radians per Gyr.  

More recently, Bailin and Steinmetz (2004, hereafter BS04) compared the orientation of the major axis determined
from the central sphere (radius of 0.6 $\times$ virial radius) of 317 halos
selected from a high resolution simulation over five time intervals (a total
period of just over 1\,Gyr).  They found that 278 (88 per cent) of their halos
did indeed rotate smoothly, with pattern speeds of around 0.15 $h$ (where  H\MySub{0} = 100 $h$ km s$^{-1}$ Mpc$^{-1}$ ) radians per Gyr. 
They noted that the pattern speeds observed in the inner regions of the undisturbed halos in their sample were not sufficient to cause the spiral patterns as discussed by \cite{bib:bureau99}. 

We note that the simulations of \cite{bib:masset03} are 2-D and the code
used was usually applied to disks around stars. When one considers the
3-D simulations of \cite{bib:bekki02}, in their figure 2f, some spiral structure is evident after
1\,Gyr for much smaller pattern speeds than those considered by
\cite{bib:masset03}. The slowest figure rotation in \cite{bib:bekki02} is 0.77
km\,s$^{-1}$kpc$^{-1}$, which is significantly larger than typical values found in BS04, but it is
unclear what would be expected in galaxies with slower pattern speeds over much longer periods of time. Here we extend the analysis of BS04 to
investigate whether one would expect coherent figure rotation over
periods much longer than the 1\,Gyr period they used.   

We have simulated the evolution of structure in a (50 $h$$^{-1}$ Mpc)$^3$
region of space, in a $\Lambda$CDM cosmology, using the N-body code GADGET \citep{bib:gadget} to evolve the
positions and velocities of 256$^3$ collisionless particles. Halos are
identified from the simulation using a friends-of-friends algorithm.
To explore figure rotation, we have identified halos from simulation outputs
at a redshift of zero and then traced them back through time. We then
determine the principal axes of each halo using the inertia tensor. By
following the motion of the principal axes of each halo over several time
steps, we are able to measure the pattern speed of the halo. Our approach
is similar to that of BS04 but we extend their analysis to test some of their assumptions.
% Firstly, in calculating the principal axes, we compare results obtained when one uses all the particles in a group identified by the group-finder, with those obtained when limiting analysis to particles in a central sphere. Secondly, we investigate using a  ``standard'' inertia tensor as well as the modified form which artificially weights central particles more heavily. We also include a discussion of the environment of the halos selected, as compared with all halos in the simulation. 

The paper is laid out as follows. In section 2, the simulations and analysis
procedure are described. In section 3, we present the results of the analysis:
the mass function, the spin parameter distribution, the outcome of halo tracing and the accretion constraints applied, the outcome of substructure elimination, the environment of halos selected for study, the
figure rotation measurements, the alignment between rotation axis and the
minor axis and, finally, the correlation between the pattern speed and halo
properties such as halo mass. In section 4, we discuss the results. \\

\section{Methodology}

\subsection{Initial conditions}
We assume the standard model for structure formation: that density
fluctuations are gaussian and the initial power spectrum is given by a power law, P$_0 \propto
$ k$^n$, where n $\sim$ 1.  As the universe moves from a radiation-dominated
to a matter-dominated phase and fluctuations begin to grow, the power spectrum becomes modified in a way that depends on the density of dark matter.  \cite{bib:bardeen} give the transfer function T (where P(k) = P$_0$(k) T) for a universe that is dominated by cold dark matter.  The normalisation of the spectrum can be obtained from observations of the Cosmic Microwave Background or from the measured abundance of clusters of galaxies.  To set initial particle positions and velocities, we used the COSMICS package \citep{bib:bertschinger96} with $\Omega_{m}$=0.3, $\Omega_{\Lambda}$=0.7 and $\sigma_8$=0.9.\\

\subsection{Simulations}
\label{sims}

We ran the publicly available parallel version of GADGET on a cluster of eight 2 GByte machines.  Doing so allowed us to follow the
evolution of $256^3$ particles.  We have chosen to simulate a cube of length
50 $h^{-1}$ Mpc.  This is sufficiently large to produce reliable periodic
simulations \citep{bib:power03}.  The choice of particle number and volume
mean that in this simulation each dark matter particle has a mass of $\sim
6 \,\times$ 10$^{8}\, \mbox{M}_\odot$.  Guided by studies of
\cite{bib:forcesoft} we have used a force softening of 30 $h^{-1}$ kpc to prevent numerical instability.  The parameters chosen for our simulation are summarised in Table \ref{sparams}. 
\ \\
\begin{table} 
\begin{center}
\begin{tabular}{|c||c|} 
Parameter & Parameter Value \\
\hline 
N$_{\mbox{particles}}$  & 256$^{3}$\\
Box Length  & $50 \, h^{-1}$ Mpc\\
M$_{\mbox{particle}}$ & $6.2 \times 10^8 \, h^{-1}$ M$_\odot$\\
Force Softening & $30 \, h^{-1}$ kpc\\
Initial Redshift & 55 \\
\hline 
\end{tabular}
\caption{\label{sparams}Simulation parameters}

\end{center}
\end{table}

\subsection{Analysis procedure}
We have identified halos (groups of ten or more particles) in the redshift
zero snapshot using FoF, a simple friends-of-friends group finder for N-body
simulations developed by the simulation group at the University of
Washington.  We have used the common convention of setting our linking length
to be 0.2 times the mean inter-particle spacing. We obtained results for two
different definitions of a halo. Firstly, we used all particles identified by
the group finder and relied on our technique for substructure elimination to
exclude groups which contained 'sub-halos'. We also tried the halo definition
of BS04, which selects a central sphere of particles within a group. Once we
have defined which particles are in a halo, we use the positions of these
particles to calculate the inertia tensor, and hence the principal axes, of
the halo of interest at each time step.

\subsubsection{Calculating the inertia tensor}
\label{sec:calcIT}
For figure rotation analysis, we model each halo as a rigid body, and
calculate the inertia tensor using

\begin{center}
\begin{eqnarray}
\label{TheIT}
 I_{ij} = \sum_{\alpha} m_\alpha \left[ \delta_{ij} \sum_k x_{\alpha,k}^2 - x_{\alpha,i}x_{\alpha,j}\right]. \nonumber
\end{eqnarray}
\end{center}
\ \\
The eigenvalues and eigenvectors of this tensor are then extracted using
Jacobi Transformations.  We have also performed the figure rotation analysis using the modified inertia tensor suggested by \cite{bib:bailin}, which artificially weights central particles in halos.

\subsubsection{Tracing halos through time and accretion constraints}
\label{mergertrees}

To follow halo properties with redshift, we need to associate each halo with its progenitor halos at a previous
timestep.  This is done with merger trees using code developed by \cite{bib:kauffmann}.  The algorithm begins at the first output
time, $z_i$, at which a halo (of 10 or more particles) can be found.  It then steps through all halos identified at the next output time, $z_{i+1}$, searching for progenitor halos in the $z_i$ output.
A halo at redshift $z_i$ is said to be a progenitor of a halo at redshift $z_{i+1}$ if, and only if, two conditions are satisfied:  
\begin{enumerate}
\item A progenitor halo at $z_i$ must contribute at least half of its particles to the halo at $z_{i+1}$ and
\item the central (most gravitationally bound) particle of
the progenitor halo at $z_i$ must be contained within the halo at $z_{i+1}$
\end{enumerate}

We then step through halos in the $z_{i+2}$ output, searching for progenitors in the $z_{i+1}$ timestep. This continues until the output time of z=0 is reached.  

Once a merger tree has been established, we exclude halos which have accreted
a significant fraction of their mass over the time period considered. In the
''halo matching''  technique of BS04, they excluded halos which accreted more
than 15 per cent of their mass over a 1 $h^{-1}$ Gyr period; that is, to be
considered for figure rotation analysis, a halo at z=0 must contain at least
85 per cent of the particles contained by the largest progenitor halo identified in the timestep which is 1 $h^{-1}$ Gyr earlier. 

%As a further constraint we follow the "Halo Matching'' technique described by BS04.  Here a progenitor halo is defined to be the halo that contributes $\ge$ 90 $\%$ of the final halo mass, they impose the additional constraint that the mass contributed to the final halo be $\ge$ 90 $\%$ of the progenitors mass.  Realising that over a long time period a halo will typically accrete a greater fraction of its mass, BSO4 relax this limit to 85 $\%$ over 1 $h^{-1}$ Gyr.  

Since we extend our
  analysis to consider halos over a much longer period of time, we have
  explored the effects of relaxing this limit over the 3 and 5
  $h^{-1}$ Gyr period and include halos which accrete as much as 
  25 per cent of their mass over 3 $h^{-1}$ Gyr and 30 per cent over 5 $h^{-1}$ Gyr.  We then consider the
  consequences of relaxing the constraint even further in our 5 $h^{-1}$ Gyr
  analysis: instead of comparing particles in a halo at z=0 with the largest
  progenitor halo identified 5 $h^{-1}$ Gyr before, we impose a constraint on
  progenitor halos at each successive timestep. Thus, at each timestep, halos
  must contain at least 70 per cent of the particles contained by the largest progenitor halo in the previous timestep. 

\subsubsection{Measuring the figure rotation of a halo}
\ \\
To determine the figure rotation of the halo we then followed a plane fitting
approach suggested by \cite{bib:bailin}, where a detailed description of the method is given.  Briefly, the method involves fitting a plane
to the principal axes of the halo of interest determined over several time steps and then solving for the
plane, {\it z = ax + by}, that best fits these axes.  Once the plane has been
determined, measuring the amount of rotation in the plane is straight forward.
The principal axis is projected onto the plane at each time step and the angle
between the projected principal axes over a time interval gives the figure
rotation of the halo.  We then used linear regression to find the best-fitting linear relation for this figure rotation, the slope giving the pattern speed of the halo. 

We have also measured the figure rotation of several halos using the method described by \cite{bib:dubinski} and found that these results are similar to the ones we obtain using the plane fitting approach.  

\section{Results}
\label{results}

Many of the results from this section are summarized in Table 2. 

\subsection{Mass function}

\begin{figure}
\begin{center}

\includegraphics[width=6cm,height=6cm,,angle=-90,keepaspectratio]{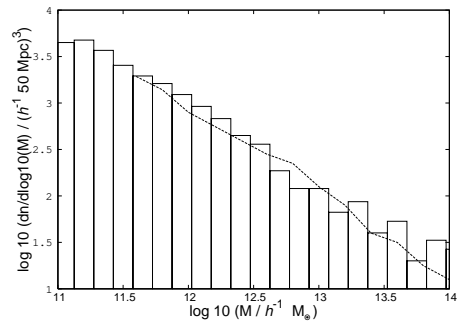}
\end{center}
\begin{MyCaption}
\caption{\label{Mymassfunction}The mass function. The histogram shows the number of halos as a function of the halo mass in the simulated region at redshift zero and the line shows the fit obtained by Jenkins et al. (2001).}
\end{MyCaption}
\end{figure}

The mass function is a measure of the number of halos as a function of their
mass.  \cite{bib:jenkins} were able to predict the mass distribution of dark
halos expected in a Cold Dark Matter universe by combining the results from
several N-body simulations.  The fit obtained by \cite{bib:jenkins} shows
good agreement with the theoretical predictions of the Sheth and Tormen
formula \citep{bib:ShethTormen}.  The mass function obtained from our simulation is shown in Figure \ref{Mymassfunction}.  This figure is a log plot of the number of halos as a function of their mass in the (50 $h^{-1}$ Mpc)$^{3}$ region of our simulation, at a redshift of zero.  We have found that our mass function agrees well with that of \cite{bib:jenkins} over the range of halos simulated.
\ \\

\subsection{Angular momentum}
\begin{figure}
\begin{center}

\includegraphics[width=6cm,height=6cm,angle=-90,keepaspectratio]{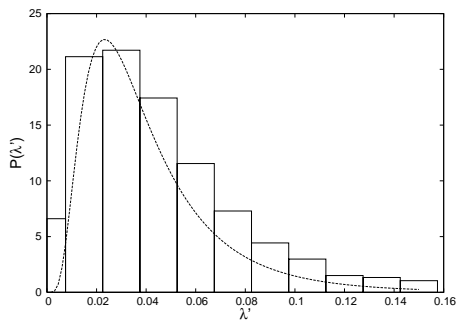}

\end{center}

\begin{MyCaption}
\caption{\label{spin}
This figure shows the distribution of the dimensionless spin parameter
$\lambda'$ for all simulated halos containing at least 200 particles, together
with the log-normal distribution given by equation \ref{lognormal} with fitted
parameters $\lambda_0' =  0.034 \,\, \pm \,\, 0.003$ and $\sigma = $ 0.62 $ \,\, \pm \,\,0.08$.}
\end{MyCaption}
\end{figure}

We use the computationally convenient expression for the spin parameter, defined by \cite{bib:angmom} as
\begin{eqnarray}
\label{easierlambda}
\lambda^{\prime} \equiv \frac{J}{\sqrt{2} \, M \, V \, R}
\end{eqnarray}
\ \\
to calculate the spin parameter of each of our simulated halos.  This
expression describes the spin parameter for a sphere of radius {\it R} (taken to be
the virial radius), where {\it M} is the mass contained within the sphere,
{\it J} is the angular momentum of the sphere, and {\it V} is the circular
velocity at the radius {\it R}.  Figure \ref{spin} shows the distribution of
the spin parameter found in our simulation.  The spin parameter was calculated
for all simulated halos which have more than 200 particles (M  $ >\, 10^{11}\,\, h^{-1}\mbox{M}_\odot$).  The curve is the log-normal distribution given by 

\begin{eqnarray}
\label{lognormal}
  P(\lambda') = \frac{1}{\lambda' \sqrt{2 \, \pi } \sigma} \, \exp \left( - \frac{\ln^2 \left( \frac{\lambda'}{\lambda_0'} \right)}{2 \,\sigma^2}\right)
\end{eqnarray}
\ \\
with best-fitting values of $\lambda_0' = $ 0.034$\,\,\pm \,\,0.003$ and $\sigma =
$ 0.62$\,\, \pm \,\,0.08$.  We have found that our best-fitting value for
$\lambda_0'$ and $\sigma$ are consistent with the values obtained by \cite{bib:barnes} and
\cite{bib:angmom}. \\

%~~~~~~~~~~~~~~~~~~~~~~~~~~~~~~~~~~~~~~~~~~~~~~~~~~~~~~~~~~~~~~~~~~~~~~~~~~~~~~
\begin{table*} 
\begin{center}
\begin{tabular}{|c||c||c||c||c|} 
& 1 $h^{-1}$Gyr & 3 $h^{-1}$Gyr & 5 $h^{-1}$Gyr & 5 $h^{-1}$Gyr relaxed\\
\hline 
Halos traced & 218 & 201 & 191 &191\\ 
Halos after accretion cut  & 183 & 62 & 13 & 74 \\
Halos after substructure cut & 115 & 38 & 5 & 48\\
Halos after BS04 substructure cut & 89 & 55 & 12 & 70\\
Coherent Figure Rotation  & 61\% & 71\% & 100\% & 73\%\\
$\mu$ (fitted) & -0.60 $\pm \,$0.03 & -0.95 $\pm \,$0.04 & -1.03 & -0.94 $\pm \,$0.02\\
$\mu$ (rad/Gyr) & 0.25h & 0.11h & 0.09h & 0.11h \\
$\sigma$ & 0.37 $\pm \,$ 0.03 & 0.27 $\pm \,$ 0.04 & & 0.21 $\pm \,$ 0.02\\
\hline 
\end{tabular}
\caption{\label{resultssum}Summary of results. Values given in the first four
  rows correspond to the number of halos, of the original 222, which remain
  after various cuts are made. The accretion cut is applied before the
  substructure cut. We give results for two different methods of doing the
  substructure cut. Values given in the last three rows are for the first
  substructure cut, not that of BS04. The fitted value of $\mu$ is also given
  in units of radians per gigayear. Only the mean is given for the 5 $h^{-1}$\,Gyr sample since it only includes 5 halos.}

\end{center}
\end{table*}
\subsection {Halo tracing and accretion constraint results}

\cite{bib:bailin} used a computationally intensive bootstrapping technique to
  show that the error in the determination of a halo's principal axes depends
  on the number of particles contained within that halo and the intrinsic
  shape of the halo.  They also showed that it was not possible to measure the
  principal axes of halos with the precision required for figure rotation
  measurements if the halos contained fewer than 4000 particles. Guided by
  their work, we identified the 222 halos in our simulation at a redshift of
  zero, which contain at least 4000 particles (that is, a mass of at least $2.5
  \times 10^{12} \, h^{-1} \, \mbox{M}_\odot $).  BS04 also showed that errors are large for halos which are almost oblate but this only affects a small number of halos in our case. We are able to trace
  218, 201, 191 halos over the  1, 3, 5 $h^{-1}$ Gyr periods respectively.

After imposing constraints on the acceptable amount of accretion over the time
period considered, as discussed in \ref{mergertrees}, we are left with samples
of 183, 62 and 13 halos over the 1, 3, 5 $h^{-1}$ Gyr periods respectively.
When we use the relaxed 5 $h^{-1}$ Gyr constraints, ie. impose a constraint at each timestep, rather than a constraint over the entire period, we obtain 74 halos over 5 $h^{-1}$ Gyr. 

\subsection{Eliminating halos with substructure}

\ \\
We are concerned with the figure rotation of undisturbed halos and,
  therefore, needed to eliminate  all halos exhibiting a significant amount of
  substructure.  Identifying substructure is somewhat subjective and we
  compared various methods. The first method involved projecting the distribution of particles along three perpendicular axes and considering each axis individually. We expect the mass distribution of an undisturbed halo to be
  smooth when considered along any axis and the presence of a secondary
  peak in this distribution thus provides evidence of substructure.
 We calculated the mass contained in bins moving out from the
  centre of the halo, where the centre is defined as the most bound particle.  If any bin contained more mass than an inner bin we compared the
  mass of that bin to the total mass of the halo. After visually inspecting
  the distributions, and some experimentation, we chose to use 25 bins and
  eliminate halos if an outer bin contained more than 10 per cent of the
  total halo mass.  Applying this to our sample of
  halos over a period of 1 $h^{-1}$\,Gyr, we eliminated 68 halos on the
  basis of substructure.  24 and 8 halos needed to be excluded from the
  sample over the 3 and 5 $h^{-1}$\,Gyr periods respectively.  For our
  larger 5 $h^{-1}$ Gyr sample, where we constrain common particles between
  successive timesteps, 26 of the 74 halos need to be eliminated. 

We also tried using the method of \cite{bib:bailin}. Their method
involves excluding halos if the calculated pattern speed varies significantly
when one changes the size of the sphere used to define a halo. The BS04
method only considers particles within the central sphere of the halo and
substructure outside of this will not cause a halo to be eliminated.  Using this approach we found that 94, 7 and 1 of the 183, 62 and 13 halos we considered would have been eliminated. For the relaxed constraint over
  5 $h^{-1}$\,Gyr, 4 of the 74 halos need to be eliminated. We note that
our method was generally more conservative over longer time periods (over the
5 $h^{-1}$\,Gyr period we rejected 35 per cent of halos while 5 per cent would have been
eliminated using their method) but there were a few halos that would have been excluded using their method and were not excluded using ours. 
%Over 1$h^{-1}$ Gyr, their method rejects a larger fraction of halos (51\%) in our simulation than in their simulation (31\%). This may be because their simulation probes lower mass halos which are less likely to be rejected. 
%Comparing the halos eliminated we find that BS04 would have eliminated 51 of
%the 115, 5 of the 38 and 1 of the 5 halos we believe to be undisturbed over the 1$h^{-1}$, 3$h^{-1}$, 5$h^{-1}$ Gyr periods respectively. 

\subsection{Mass density in surrounding environment}

\begin{figure}
\begin{center}
\begin{tabular}{cc}

\includegraphics[width=6cm,height=6cm,angle=-90,keepaspectratio]{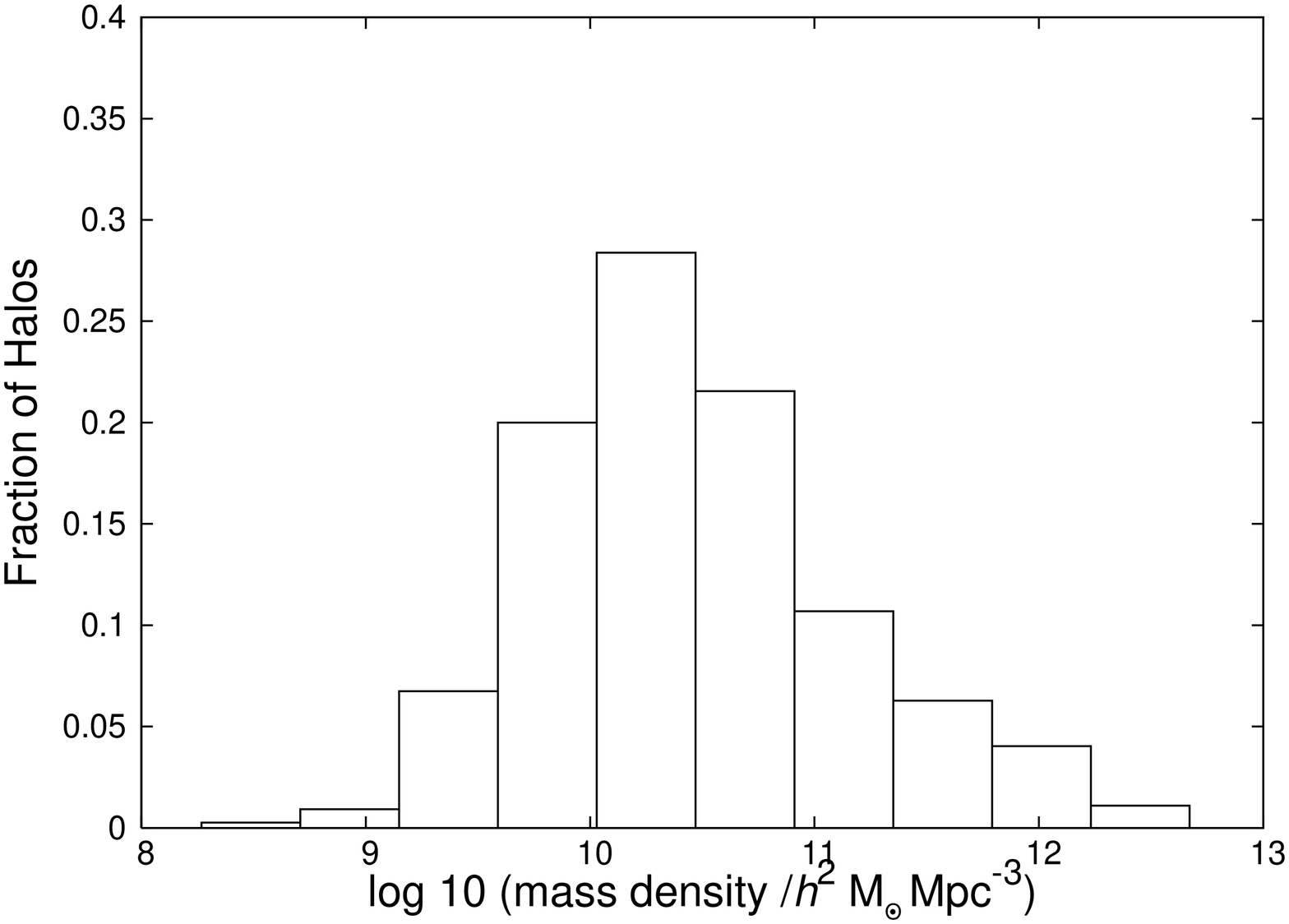}\\
\includegraphics[width=6cm,height=6cm,angle=-90,keepaspectratio]{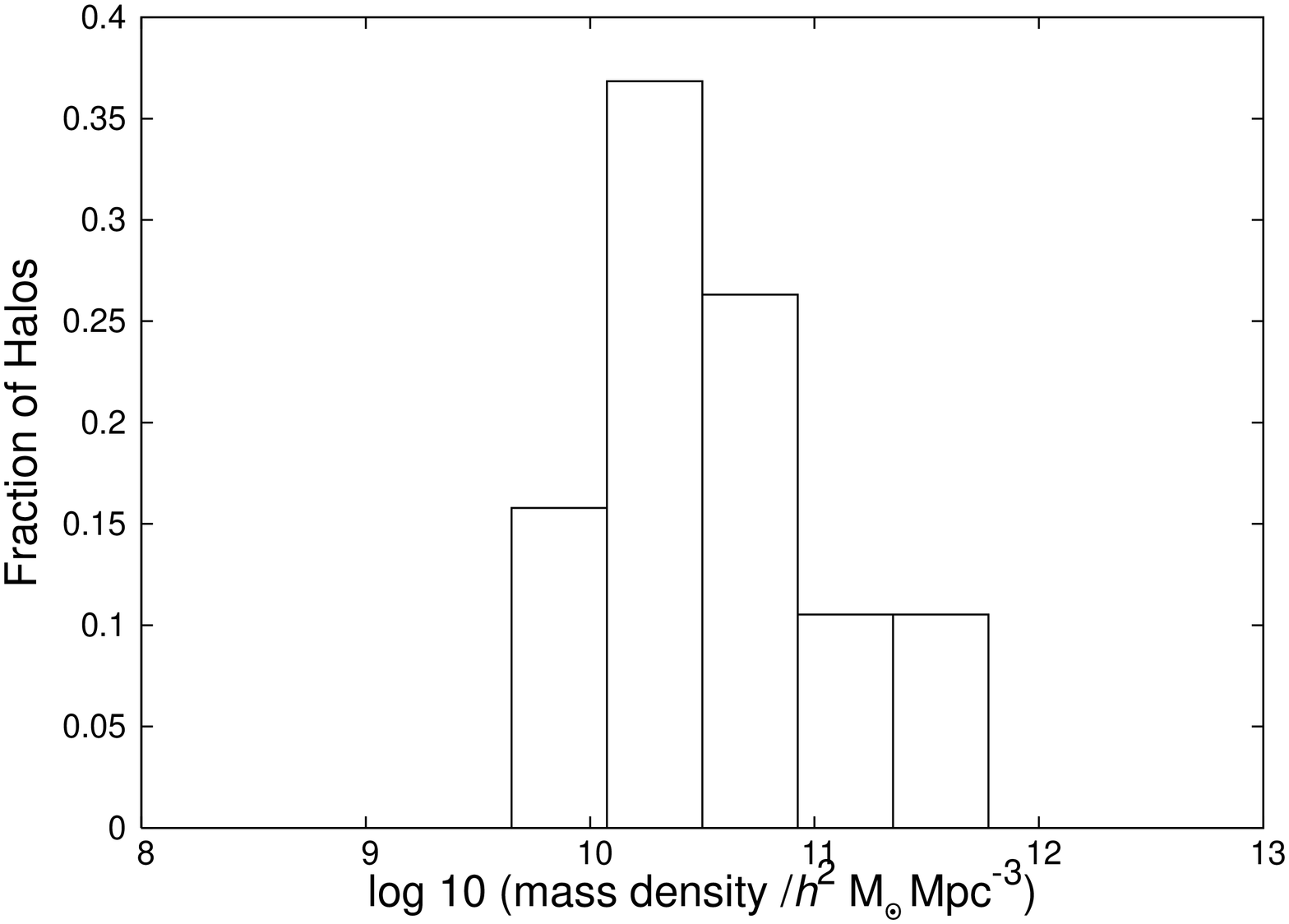}
\end{tabular}
\end{center}
%\end{figure}
\begin{MyCaption}
\caption{\label{massdens}The mass density in surrounding environment. These histograms show the fraction of halos versus the log of the mass density of the surrounding environment.  Top: The surrounding environment of all simulated halos.  Bottom:  The environment of the halos selected for 3\,Gyr analysis.}
\end{MyCaption}
\end{figure}

In order to explore the environment of our halos, we have considered the mass
density of the area surrounding the halos at z=0.  To do this we have, for
each halo, found the number of particles within 5 $h^{-1}$ Mpc of the halo.
The sum of the surrounding mass is then divided by the volume of a sphere with
radius of 5 $h^{-1}$ Mpc.  The results are shown in Figure \ref{massdens}
where we plot the fraction of halos as a function of the log of their mass
density.  The top figure shows the distribution of mass density of all the
halos identified from the simulation, while the bottom histogram shows the
distribution of our sample of undisturbed halos.  While we lose a few of the most massive halos in our selection, the undisturbed halos appear to occur in an environment
which is slightly more dense than average.  This is consistent with the
results of \cite{bib:environment} who found that, for z $<$ 1, the merger rate of cluster halos to be 3 times lower than that of isolated halos and twice as low as halos that end up in groups.  With this in mind, we expect that undisturbed halos would occur in a more dense environment. 

\subsection{Figure rotation}

\begin{figure}
\begin{center}
\begin{tabular}{c}  
\includegraphics[width=6cm,height=6cm,angle=-90,keepaspectratio]{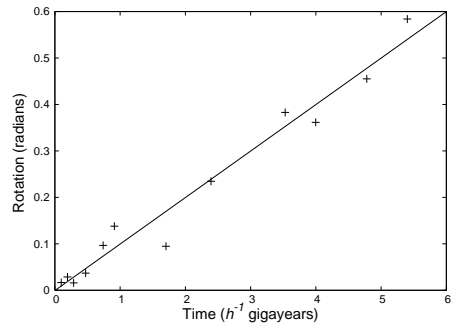} \\
\end{tabular}
\end{center}
\begin{MyCaption}
\caption{\label{patternspeed3gyr}Pattern speed of a single halo measured over
  5 $h^{-1}$\,Gyr.  Each point represents the angle (in radians) that the figure has rotated over the time interval given.   The slope of this plot (obtained by linear regression) gives the pattern speed of the halo.}
\end{MyCaption}
\end{figure}

To determine the pattern speed of a halo, we have followed each halo over a
period of 5 $h^{-1}$\,Gyr (or 1 or 3 $h^{-1}$\,Gyr if the halo does not satisfy selection criteria over longer periods).  Using linear regression, we have found the best-fitting linear relation for the figure rotation of the halos as a function of
time.  The pattern speed of the halo is given by the slope of the linear fit.
We have calculated the one-sigma limit of the slope, and we have taken this to
be the error in the pattern speed. Figure \ref{patternspeed3gyr} shows the
figure rotation of one of our halos that has remained unaffected by substructure over
a 5 $h^{-1}$\,Gyr period. Results are similar for other halos. A linear fit is adequate over the whole period and there is no indication that the pattern speeds are systematically changing over time. 

\begin{figure}
\begin{center}
\begin{tabular}{c c}

\includegraphics[width=6cm,height=6cm,angle=-90,keepaspectratio]{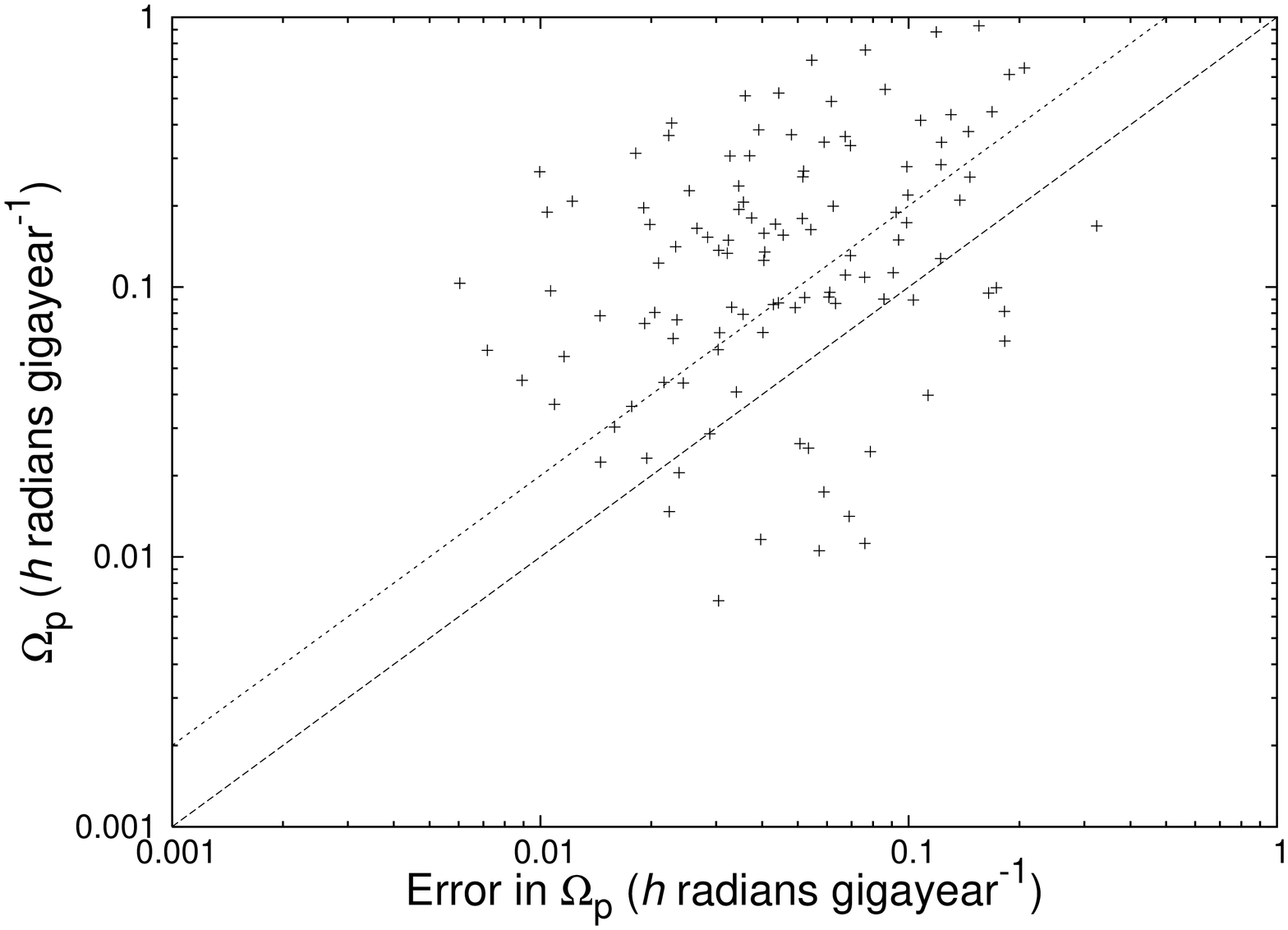}\\
\includegraphics[width=6cm,height=6cm,angle=-90,keepaspectratio]{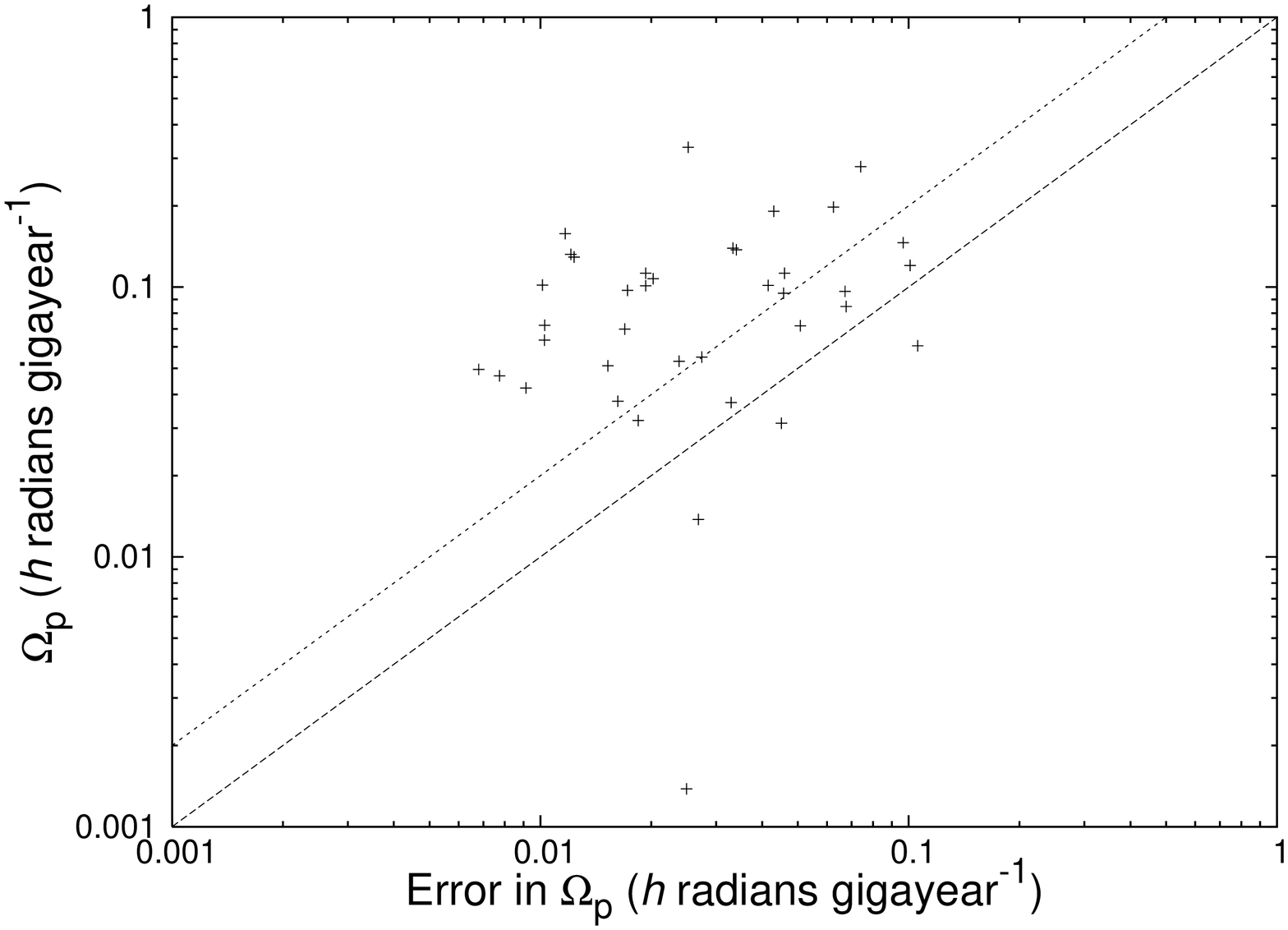}\\
\includegraphics[width=6cm,height=6cm,angle=-90,keepaspectratio]{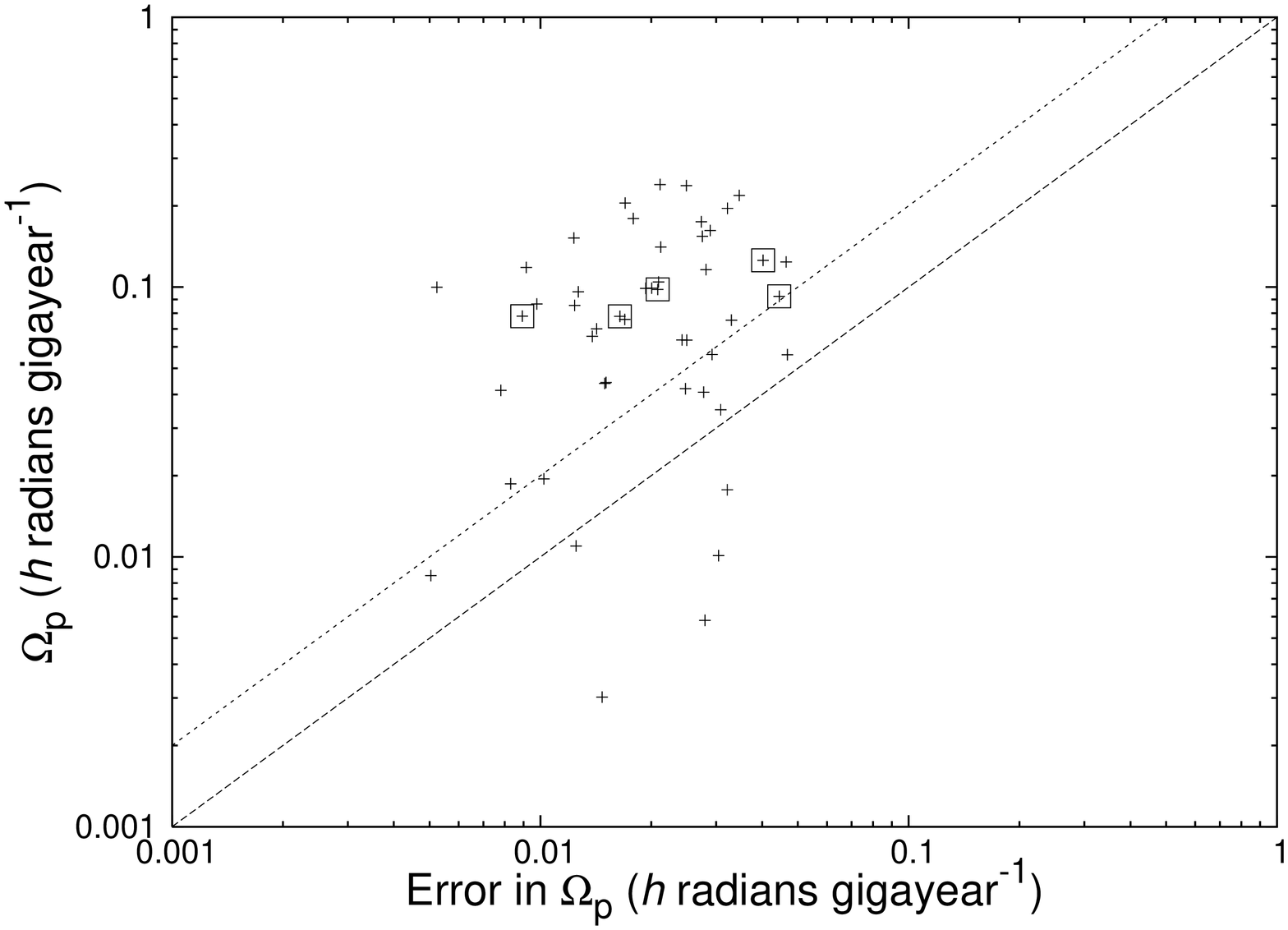}
%\\
%\includegraphics[width=6cm,height=6cm,angle=-90,keepaspectratio]{Figures/ErrorInSpeed_z13r.eps}
\end{tabular}
\end{center}
\begin{MyCaption}
\caption{\label{Error_z7}Measured pattern speeds of the simulated halos versus the estimated
  error in pattern speed.  The dashed line represents the points at which the
  pattern speed is equal to the estimated error (1 $\sigma$ limit), the dotted
  line represents the points at which the pattern speed is equal to twice the
  error.  Halos with pattern speeds less than the twice the error are not
  considered to be rotating coherently. Top: Halos considered over 1 $h^{-1}$\,Gyr
  period.  Centre: Halos considered over 3$h^{-1}$\,Gyr. Bottom: Halos considered
  over 5 $h^{-1}$\,Gyr period using relaxed accretion constraints. Boxed crosses
  mark the pattern speeds and errors of the 5 halos that meet the stricter accretion criteria similar to that used in the top and middle plots and in BS04).}
\end{MyCaption}
\end{figure}

\begin{figure}
\begin{center}
\begin{tabular}{cc}
  \includegraphics[width=6cm,height=6cm,angle=-90,keepaspectratio]{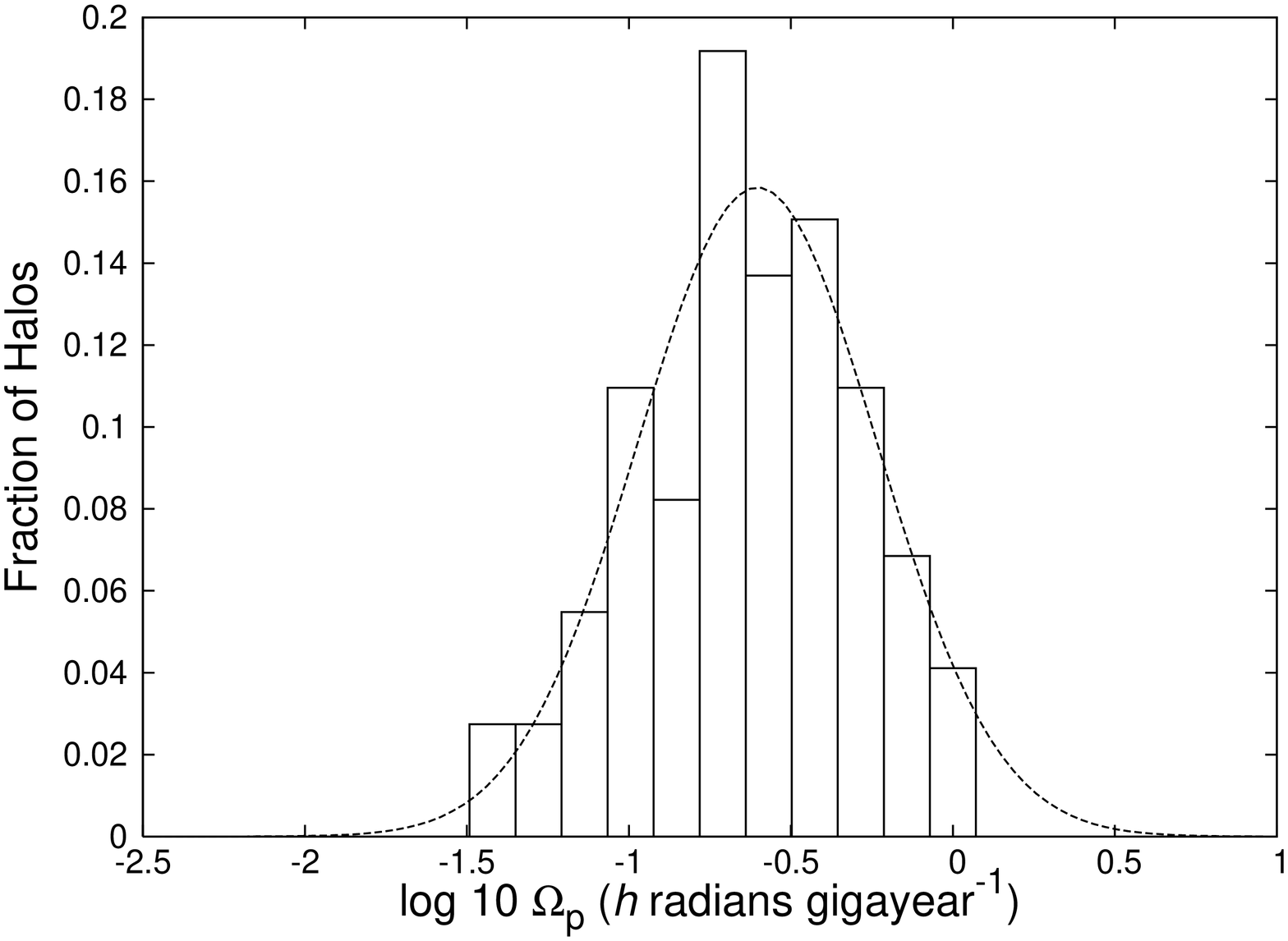} \\
\includegraphics[width=6cm,height=6cm,angle=-90,keepaspectratio]{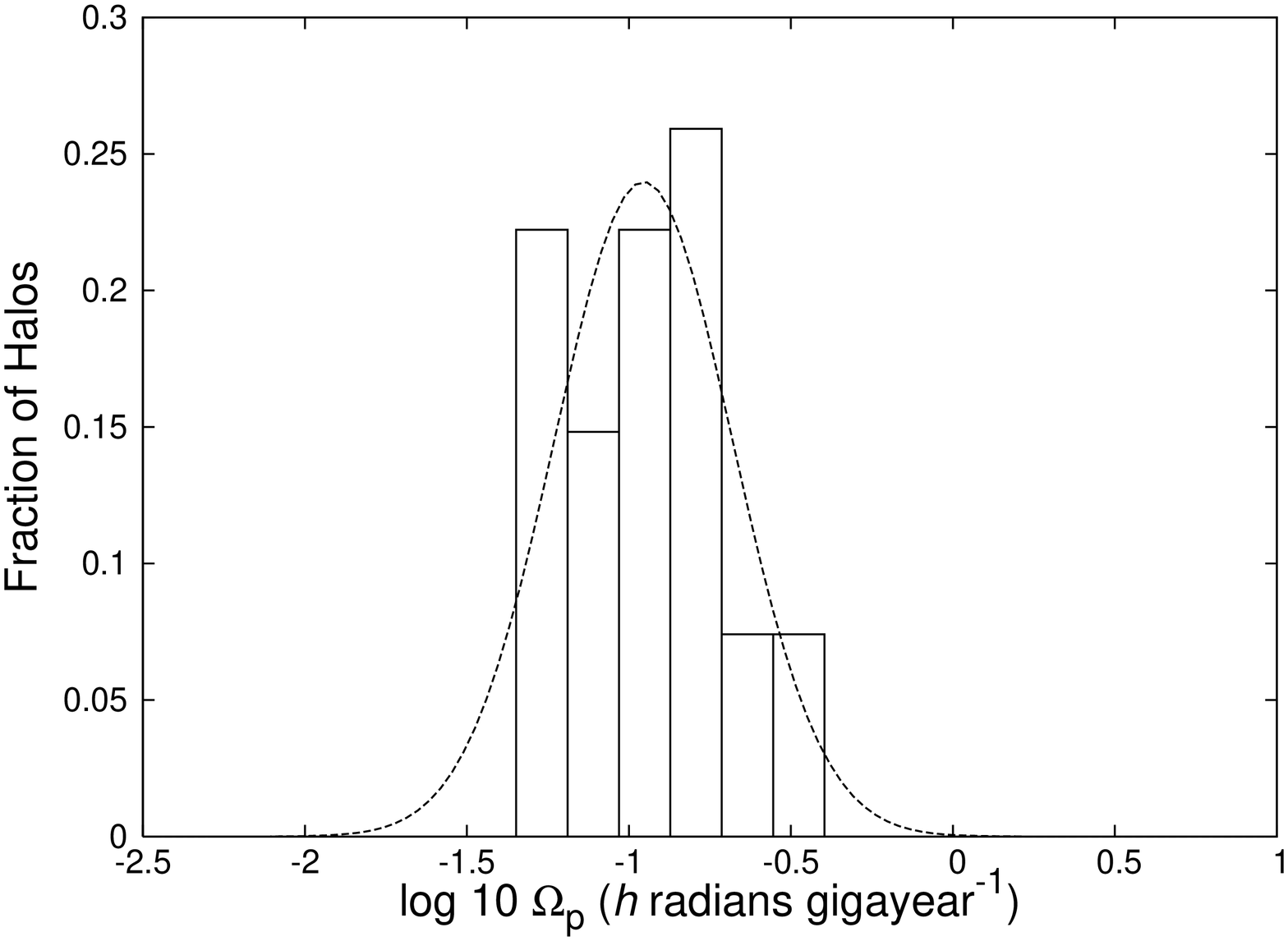}\\
\includegraphics[width=6cm,height=6cm,angle=-90,keepaspectratio]{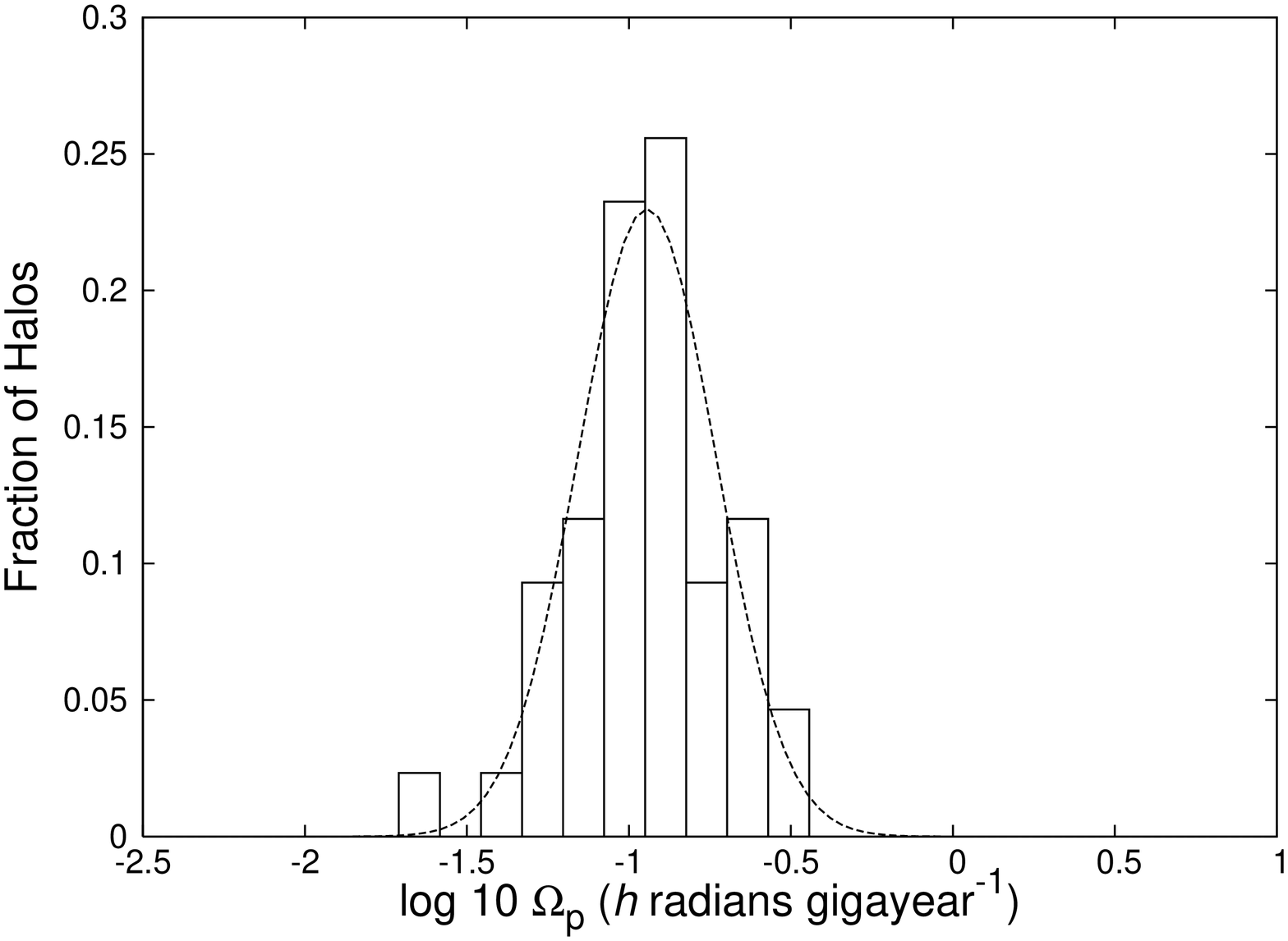}
\end{tabular}
\end{center}
\begin{MyCaption}
\caption{\label{Npatterndist3gyr}Pattern speed distribution for rotating
 halos.  This histogram shows the distribution of the pattern speeds
 calculated from our undisturbed halos.  The line is the log-normal
 distribution (equation \ref{logpattern}).  Top: Halos Rotating over 1
 $h^{-1}$\,Gyr (best-fitting values of $\mu = -0.60 \pm 0.03$ and $\sigma =
 0.37 \pm
 0.03$).  Centre: Halos
 Rotating over 3 $h^{-1}$\,Gyr ($\mu = -0.95 \pm 0.04$ and $\sigma = 0.27 \pm
 0.04)$.    
Bottom: Halos Rotating over 5 $h^{-1}$\,Gyr (relaxed constraints) ($\mu = -0.94 \pm 0.02$ and $\sigma =
0.21 \pm  0.02)$.}
\end{MyCaption}
\end{figure}

Figure \ref{Error_z7} shows the pattern speeds obtained for our halos.  The
115 halos observed over 1 $h^{-1}$\,Gyr are shown in the top plot, the middle
plot shows the 38
halos observed over 3 $h^{-1}$\,Gyr, the bottom plot shows the halos considered
  to be undisturbed over 5 $h^{-1}$\,Gyr period using relaxed accretion constraints (boxed crosses
  mark the pattern speeds and errors of the 5 halos that meet the stricter accretion criteria similar to that used in the top and middle plots and in BS04).  The x-axis is the estimated error in pattern speeds.  The dashed line represents
the points at which the observed pattern speed is equal to the estimated error.
We have also shown the points at which the pattern speed is equal to twice the
estimated error (dotted line). Using a 2$\sigma$ cutoff for halos to be
considered rotating coherently, 63 per cent of the halos in the 1 $h^{-1}$\,Gyr
sample, 71 per cent of the halos in the 3 $h^{-1}$\,Gyr and 100(73) per cent of the halos in
the 5 $h^{-1}$\,Gyr(relaxed) sample exhibit coherent figure rotation.\\
\ \\  
We found that the distribution of pattern speeds of the undisturbed halos from our simulation was fairly well fit by the log-normal distribution 
\begin{eqnarray}
\label{logpattern}
P(\Omega_p) = \frac{1}{\Omega_p \, \sigma \, \sqrt{2\,\pi}}\, \exp\left( \frac{-\log^2\left( \frac{\Omega_p}{\mu}\right)}{2\,\sigma^2}\right) \,.
\end{eqnarray}
\ \\
Figure \ref{Npatterndist3gyr} shows the distribution of pattern speeds for the
halos considered, together with the log-normal distribution (equation
\ref{logpattern}).  Best-fitting values to this curve are given in Table 2.
%were found to be $\mu = -0.67
%\,\, \pm \,\,0.01 $, $\sigma = 0.40\,\,\pm 0.01\,\,$ over 1$h^{-1}$\,Gyr, $\mu = -1.02 \,\, \pm 0.02\,\, $, $\sigma = 0.28 \,\,\pm
%0.02\,\,$ over 3$h^{-1}$\,Gyr,  $\mu =  -1.04\,\, \pm 0.01\,\, $ and
%$\sigma = 0.04 \,\,\pm 0.01
%\,\,$ over 5$h^{-1}$\,Gyr and $\mu =  -0.98\,\, \pm 0.01\,\, $ and
%$\sigma = 0.25 \,\,\pm 0.01
%\,\,$ over the relaxed constraints 5$h^{-1}$\,Gyr.
For the undisturbed halos over 5 $h^{-1}$\,Gyr the fastest figure rotation
detected was 0.13 $h$ radians per Gyr and the average pattern speed of these
halos was 0.09 $h$ radians per Gyr. The maximum figure rotation for the halos
with relaxed contraints over 5 $h^{-1}$\, is 0.24 $h$ radians per Gyr. Adjusting our cutoff to include only those halos with pattern speeds greater
than three times the error, does not significantly change the pattern speed
distribution.

\subsection{Alignment between figure rotation axis and minor axis}

\begin{figure}
\begin{center}

\includegraphics[width=6cm,height=6cm,angle=-90,keepaspectratio]{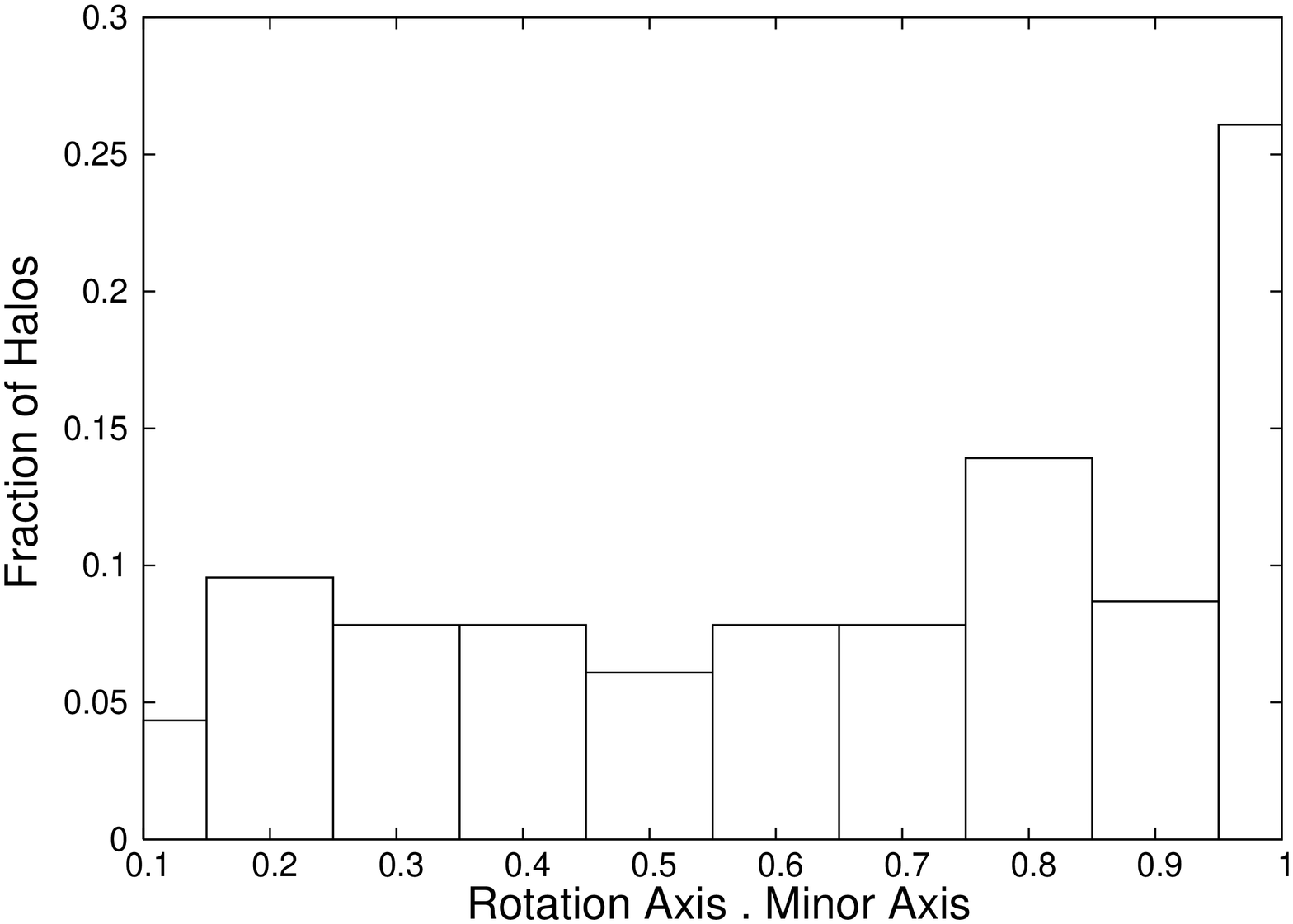}\\
\includegraphics[width=6cm,height=6cm,angle=-90,keepaspectratio]{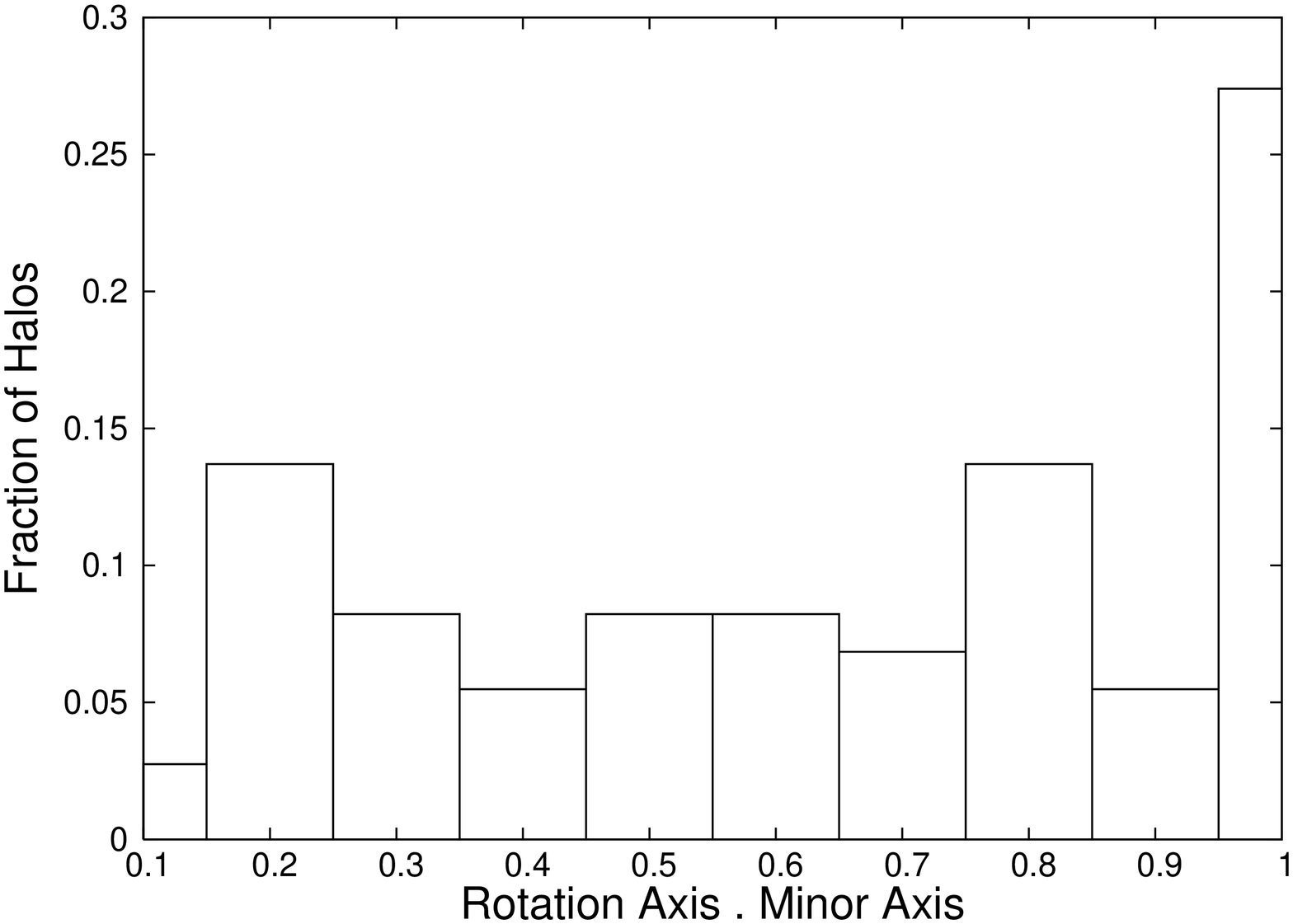}\\
\includegraphics[width=6cm,height=6cm,angle=-90,keepaspectratio]{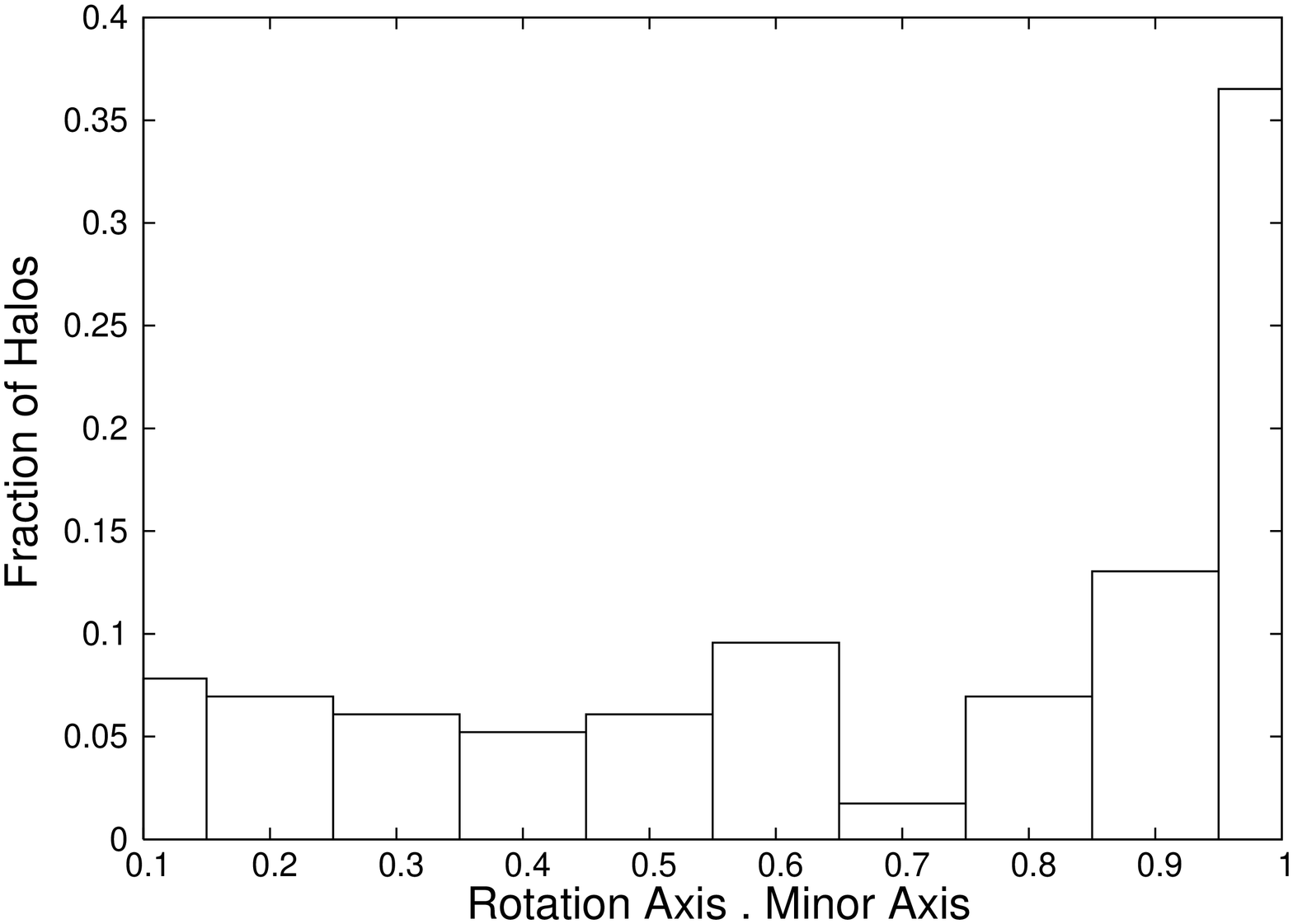}
\end{center}
\begin{MyCaption}
\caption{\label{rotation.minor}Alignment between figure rotation axis and minor axis. 
This figure shows the dot product of the figure rotation axis and the minor
axis.  Top: axes calculated for the 115 halos in our sample using the whole
halo and the 'standard' inertia tensor.  Middle: axes calculated for the 73
halos found to undergo coherent rotation.  Bottom: axes (for 115 halos) calculated using a central sphere of 0.6 times the virial radius and a modified inertia tensor.}
\end{MyCaption}
%\end{center}
\end{figure}
\ \\
\cite{bib:dubinski}, \cite{bib:pfitzner99} and \cite{bib:bailin} all noted
that for the majority of their simulated halos the major axis appeared to
rotate around the minor axis.  To explore this, we have determined the dot
product between the figure rotation axis and the minor axis of each of the halos in
our sample.  Figure \ref{rotation.minor} shows our results when we use a standard inertia tensor and do not limit ourselves to central particles in the definition of the halo.  It appears that
these axes are well-aligned for at least a third of the halos.  We note, however, that the fraction of halos exhibiting this alignment is smaller than in other studies.  If we consider only the
spherical centre of the halos, the percentage of halos exhibiting this
alignment increases to $\sim$ 50 per cent which is more consistent with the 58
per cent found by \cite{bib:bailin}. Using a modified inertia tensor also slightly increases the fraction of halos found to be rotating around the minor axis.

\subsection{Effect of halo properties on pattern speed}

\begin{figure}
\begin{center}
\begin{tabular}{c c}

  \\
%  \includegraphics[width=6cm,height=6cm,angle=-90,keepaspectratio]{Figures/NewPatternSpeedvsMass.eps}\\
 % \includegraphics[width=6cm,height=6cm,angle=-90,keepaspectratio]{Figures/PSvsFrac1.eps}\\
%  \includegraphics[width=6cm,height=6cm,angle=-90,keepaspectratio]{Figures/PSvsFrac2.eps}
%Also want frac vs pattern speed
\includegraphics[width=6cm,height=6cm,angle=-90,keepaspectratio]{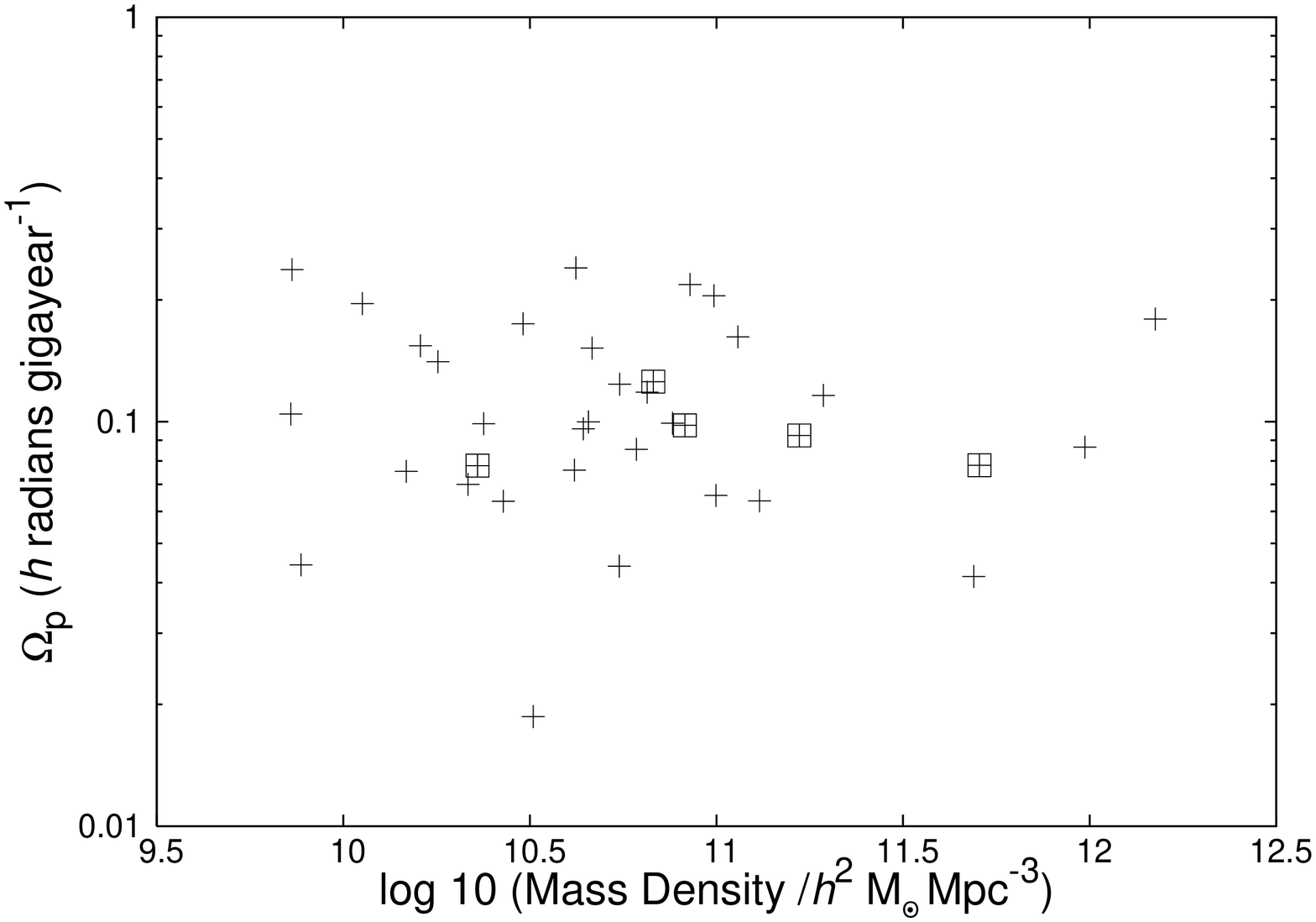}\\
\includegraphics[width=6cm,height=6cm,angle=-90,keepaspectratio]{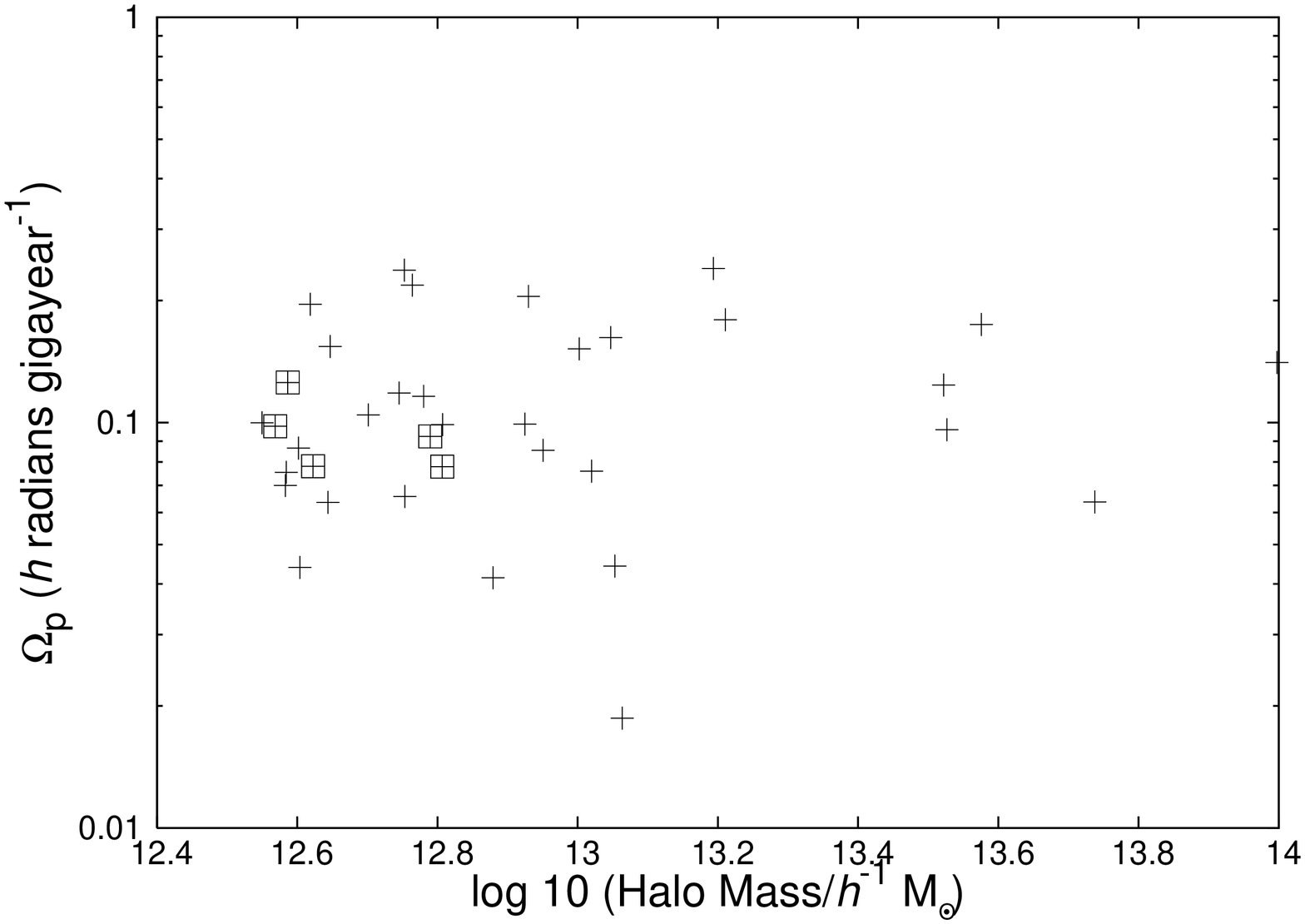}\\
 \includegraphics[width=6cm,height=6cm,angle=-90,keepaspectratio]{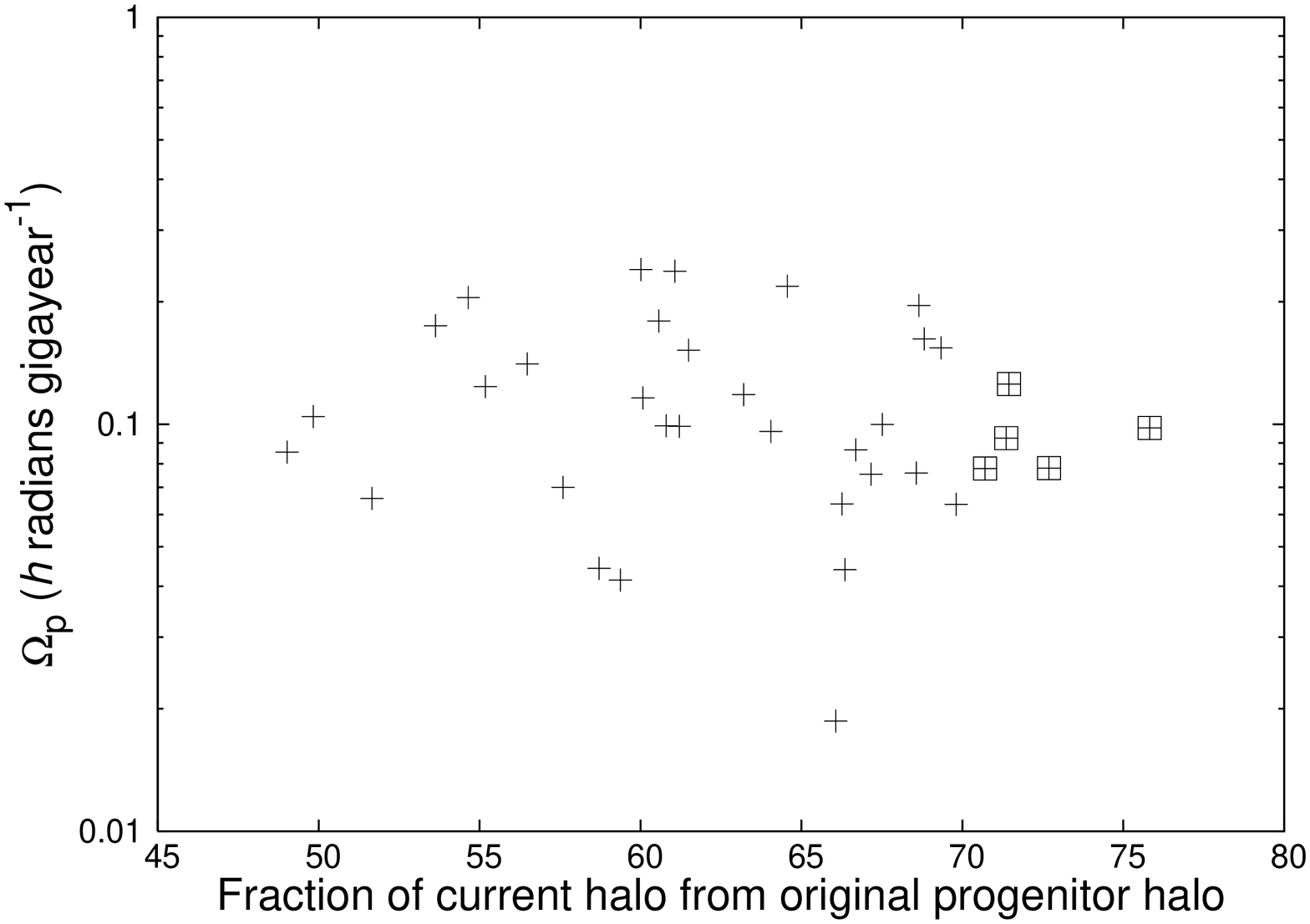}
\end{tabular}
\end{center}
\begin{MyCaption}
\caption{\label{PatternSpeedvsHalo}Pattern speed versus halo properties.
We have compared pattern speeds to (from top to bottom) the mass density in the surrounding
environment, the halo mass and the fraction of current halo particles present in progenitor halo.  The pattern
speeds are calculated over the 5 $h^{-1}$\,Gyr period.  The halo mass and mass
density in surrounding environment are determined at z=0. Boxed crosses are halos which meet the stricter accretion criteria of BS04.}
\end{MyCaption}
\end{figure}

We have investigated the correlation of pattern speed with halo properties
such as mass, density of the surrounding environment and the amount of matter
accreted over 5 $h^{-1}$ Gyr (fraction of particles
common to progenitor and current halos).   We plot results in Figure \ref{PatternSpeedvsHalo}, demonstrating that pattern speed is not correlated with any of these halo properties. The final plot demonstrates that our results are not sensitive to the the accretion cut values we use.

%~~~~~~~~~~~~~~~~~~~~~~~~~~~~~~~~~~~~~~~~~~~~~~~~~~~~~~~~~~~~~~~~~~~~~~~~~~~~~~
\section{Discussion and conclusions}
\label{discussion}
\ \\
We have found that the mass function determined from our simulation is well
fit by the mass function obtained by \cite{bib:jenkins} using N-body
simulations.  The distribution of the spin parameter found from our simulation was well fit by a log-normal distribution (equation \ref{lognormal}) with a best-fitting value of $\lambda_0' $ = 0.034.  This agrees well with the values found by \cite{bib:barnes} and \cite{bib:angmom}.  Both of these tests suggest that the simulations we have run are reliable, agreeing with many previous simulations as well as with theoretical predictions.

We selected halos which had not accreted a large fraction of their material and which did not show evidence for significant substructure and found that 63 per cent of the 115 halos considered over 1 $h^{-1}$\,Gyr did exhibit
coherent figure rotation, where `coherent' is arbitrarily defined to be pattern speeds greater than two times the
estimated error.  The distribution of the pattern speeds over 1 $h^{-1}$\,Gyr was well-fit by a log-normal distribution centred at log$(\Omega_p) = -0.60 \pm 0.01$
($\Omega_p$ = 0.24 $h$ radians per Gyr).  The pattern speeds of these
halos, determined over a 1 $h^{-1}$\,Gyr period, are consistent with 
the lower end of the 0.1 - 1.6 radians per Gyr found by \cite{bib:dubinski}, but slower than the 1.1
radians per Gyr detected by \cite{bib:pfitzner99}. Our distribution of pattern  speeds peaks at slightly higher values than those found by \cite{bib:bailin}, but the difference is not very significant. It appears that limiting oneself to the central sphere of a group and using a modified inertia tensor do not significantly influence the results. The accretion cut and the substructure cut affect the number of halos considered, but the pattern speed distributions are not very sensitive to the details.

%When we studied halos without substructure over longer periods, 71\% were found to have 
%coherent figure rotation over a 3$h^{-1}$\,Gyr period and 100\%
%over a 5$h^{-1}$\,Gyr period.  \\

%Again, the distributions were well fit by a log-normal
%curve, centred at log$(\Omega_p) = $ -0.98 $ \pm \,\, 0.02$ ($\Omega_p $ =
%0.10$h$ radians per Gyr) over 3$h^{-1}$\,Gyr and log$(\Omega_p) = $ -1.05 $ \pm \,\, 0.01$ ($\Omega_p $ =
%0.09$h$ radians per Gyr) over 5$h^{-1}$\,Gyr. \\

The pattern speed distributions appear to shift to lower speeds as one looks
over longer periods. The difference between the 3 $h^{-1}$\,Gyr and 5
$h^{-1}$\,Gyr distributions is not significant but the difference between the
1 $h^{-1}$\,Gyr and 5 $h^{-1}$\,Gyr distributions is significant. This result holds no matter what halo definition, accretion cut or method of substructure removal is used. Presumably, studying the halos over a
longer time period allows us to detect smaller figure rotations above the noise but, over 5 $h^{-1}$\,Gyr, there are no halos with the larger figure rotations seen over 1 $h^{-1}$\,Gyr, suggesting that accretion and/or mergers allow for increased pattern speeds. In general, the speeds found are much slower than those \cite{bib:masset03} claim are required for spiral structure in NGC2915 but, given the long periods of time over which these halos show coherent figure rotation and the results of \cite{bib:bekki02}, we suggest that further hydro-simulations would be required before strong conclusions can be drawn for all galaxies.   

The sample of undisturbed halos we have chosen for analysis are generally
found to be in more dense environments than the average halos in the
simulation.  This is consistent with the results of \cite{bib:environment} who
suggest that, at low redshifts (z$<$1), the merger rate is lower in denser
environments.  

In studying the relationship between the figure rotation axis and the minor
axis, we found alignment for many of the halos but a significant fraction
which were not very well aligned.  The two axes are within 25$^\circ$ of each
other in 30 per cent of the halos, using the ``whole-group'' definition of a halo.  If we restricted
our analysis to the central region and used the modified inertia tensor, as did \cite{bib:bailin}, our results are more similar to theirs. 

We have found no correlation between halo properties such as halo mass, or environment, and the pattern speed. 

\section*{Acknowledgements}The financial assistance of the South African National Research Foundation (NRF) towards this research is hereby acknowledged.  Opinions expressed and conclusions arrived at, are those of the author and are not necessarily to be attributed to the NRF.

\end{document}